\documentclass[%
 reprint,
superscriptaddress,
preprintnumbers,
 amsmath,amssymb,
 aps,
]{revtex4-2}

\usepackage{graphicx}% Include figure files
\usepackage{dcolumn}% Align table columns on decimal point
\usepackage{bm}% bold math
\usepackage[normalem]{ulem} 
\usepackage{tikz}
\usetikzlibrary{decorations.pathmorphing}
\usetikzlibrary{decorations.markings}
\usepackage{verbatim}

\newcommand{\ate}[1]{Y_{ #1}}

\begin{document}

\preprint{IPPP/20/62}

\title{Kick-alignment: matter asymmetry sourced dark matter}% Force line breaks with \\

\author{Rodrigo Alonso} %\email{unavailable for comment}
\affiliation{Institute for Particle Physics Phenomenology, Durham University, South Road, Durham, DH1 3LE}
\author{Jakub Scholtz} %\email{wouldntyouliketoknow}
\affiliation{Institute for Particle Physics Phenomenology, Durham University, South Road, Durham, DH1 3LE}
\affiliation{Dipartimento di Fisica, Universit\'a di Torino, via P. Giuria 1, I–10125 Torino, Italy}%Lines break automatically or can be forced with \\

\date{\today}% It is always \today, today,

\begin{abstract}

We investigate the possibility that the dark matter abundance is sourced by the baryon/lepton asymmetry of the early Universe. It turns out that a Goldstone field of a local classically preserved symmetry in the Standard Model experiences a kick during a period of baryon/lepton number generation. This mechanism can be regarded as dynamical generation of initial conditions for misalignment yet the prediction for relic abundance presents an inverse dependence on the coupling in parallel with freeze-in vs freeze-out. We explore two realizations of this mechanism and show that in conjunction with Leptogenesis, it is possible to identify a viable promising region of parameter space for dark matter production with mass $10$~MeV -- $1$~GeV and decay constant $f$ in the range of $10^{10} - 10^{12}$~GeV.
\end{abstract}

\maketitle

%\tableofcontents

\section{\label{sec:level1} Introduction}

Our Universe shows a surprising coincidence: the mass densities of its matter components, baryonic matter and dark matter (DM), are eerily similar down to a factor of four or so. This is odd because their abundance seems to be set by different, often unrelated mechanisms. The baryon abundance is set by a baryon number asymmetry, whose origin is unknown to us. On the other hand, the dark matter abundance can be set by any number of mechanisms -- freeze-out, freeze-in, misalignment, number asymmetry (shared or not shared with the Standard Model). At the end, all of the above mechanisms have enough free parameters that for a range of couplings and DM masses, they can generate just the right amount of DM.

This work will study a possible connection between asymmetry generation and dark matter production. This interplay has already been explored in the literature: the initial ideas were discussed in the 90s~\cite{Barr:1991qn,Kaplan:1991ah,Thomas:1995ze} and many more were published during a resurgence of this topic about a decade or two later, amongst those were~\cite{Kaplan:2009ag,Buckley:2010ui,Blennow:2010qp,Cheung:2011if,Cui:2020dly}. Of particular interest to us are models in which a condensate triggers baryogenesis~\cite{Cohen:1987vi,Cohen:1988kt}  while being itself the dark matter~\cite{Co:2019jts}.

In this work we will present a different explicit realization of this connection.

The one mainstay of DM models is that they should explain why we have not seen DM in any non-gravitational experimental apparatus so far. We are going to circumvent this (somewhat serious) constraint by coupling the DM to a divergence of a (mostly) conserved current of the Standard Model (SM) such as baryon number through the operator:
\begin{align}
    \mathcal L_{int}= \frac{\phi}{f} \nabla_\mu J^\mu_Q\,,
    \label{eq:operator}
\end{align}
where the symmetry and the current, to be explicit, are defined by
\begin{align}
    U(1)_{Q}:&\quad Q=\cos(\alpha)  B-\sin(\alpha)  L\,,\\
    J_Q^\mu &=-\frac{\partial \mathcal L(G\psi)}{\partial_\mu \theta} \quad G=e^{iQ\theta(x)}.
\end{align}
As a result, the DM and the baryons (or leptons) only interact when $Q$-number is being violated, such as during baryogenesis and in extremely rare events today (which we haven't observed yet).

There are two ways one can generate DM density this way. A quick change of basis, as we will see, turns the classic low dimension baryon/lepton number violating operators into interactions terms between SM and the DM:
\begin{align}
-\frac{i\phi}{f}\left(-2s_\alpha\frac{y_N^2}{2M_N}(\ell H)^2 +(c_\alpha-s_\alpha)\frac{y_X^2}{M_X}q q q \ell\right)\,
\end{align}
and thanks to these the DM production proceeds through a UV-sensitive freeze-in.

However, there is another curious way in which the SM plasma can transfer its energy into the dark sector. During lepto/baryogenesis $\nabla_\mu J^\mu_Q \neq 0$ so the $\phi$ field feels a source that, under favourable conditions does work on the $\phi$ field and deposits energy into its oscillations, hence generating dark matter density\footnote{A reader who appreciates non-generic baryogenesis models might notice that what we propose is in fact a reverse of the Spontaneous Baryogenesis}. For reasons that will become apparent we will call this mechanism the `Kick-alignment' -- the main focus of this work.

First, the dynamics of our candidate are discussed in  section \ref{sec:mechanism} which contain the description of the highlighted production mechanism in \ref{Sec:CH} and comparison with thermal production in \ref{Sec:TH}. Additional computational details can be found on the appendices  \ref{sec:matching} and \ref{App:Int}.
  Two explicit realizations for this mechanism can be found in section \ref{RealZ} and we finish with discussion and conclusions in section \ref{sec:conclusion}.

\section{The Mechanism}\label{sec:mechanism}

The coupling of eq.~(\ref{eq:operator}) 
 is indeed most naturally and familiarly realized if  dark matter is the (pseudo) Goldstone boson of the $Q$-symmetry. Let us write the action as
\begin{align}\nonumber
    S_{\rm eff}=\int \sqrt{|g|}d^4x\Big(&\frac{1}{2}(\partial\phi)^2-m^2f^2\cos\left(\frac{\phi}{f}\right)-\frac{\partial_\mu \phi}{f} J^\mu_Q\\&+\mathcal L_{\rm Q}+\mathcal L_{Q\!\!\!\backslash }, \Big)\label{ACT}
\end{align}
where $m$ is the mass and $f$ the decay constant that gives the periodicity of the field $\phi+2\pi f$
and  the last piece is the effective action that violates this symmetry. In its absence the coupling of ordinary matter to $\phi$ vanishes since so does $\partial_\mu J^\mu_Q$. This can be seen most explicitly by the field dependent transformation:
\begin{align}\label{Rotphi}
    G(\phi)&=e^{-i Q\phi/f}\\
    \delta \mathcal L_{ {\rm Q}}(G\psi)&=\frac{\partial_\mu\phi}{f} J_Q^\mu,\nonumber
\end{align}
which cancels out the derivative coupling in the first line of eq.~(\ref{ACT}).
%while couplings to $\phi$ shift to $\mathcal L_{Q\!\!\!\backslash} (G(\phi)\psi)$. 
To leading order in $1/f$, the non-derivative fermion operators turn into:
\begin{align}\label{eq:psiQ}
    \psi_1 \ldots \psi_n \rightarrow \psi_1 \ldots \psi_n - \frac{i \phi}{f}\left(\sum_i Q_i \right)\psi_1 \ldots \psi_n,
\end{align}
which makes it obvious only explicitly $Q$-violating operators lead to non-zero couplings with $\phi$.
It is important to remark that, unless we consider an exact symmetry in the the SM (e.g. $Q=B-L$), there is a $Q\!\!\!\backslash$ contribution in the effective action from SM physics. 

The case in which this violating term is given by the SM Sphaleron processes is an illustrative example. The zero temperature contribution is very much suppressed by the non-perturbative nature of the effect. Whereas at high temperature, the violation is thermal and unpressed. This allows for its production at high energy while simultaneously it leads to a very weak coupling today and hence cosmic scale stability of the DM.

 Similar consequences follow if one considers $Q=L$ and the Seesaw model, that is, the Majoron: at high energy the mass term of heavy RH neutrinos breaks $L$ whereas at low energies the breaking and coupling comes instead from the tiny LH $\nu$ masses~\cite{Chikashige:1980ui,Gelmini:1980re}. Our scenario will indeed present parallels with the Majoron as will be made explicit.

%%%%%%%%%%%%%%%%%%%%%%%%%%%%%%%%%%%%%%%%%%%%%%%%%%%%%%%%%%
\subsection{Kick-alignment}\label{Sec:CH}
\begin{figure}
    \centering
    \includegraphics[width=.45\textwidth]{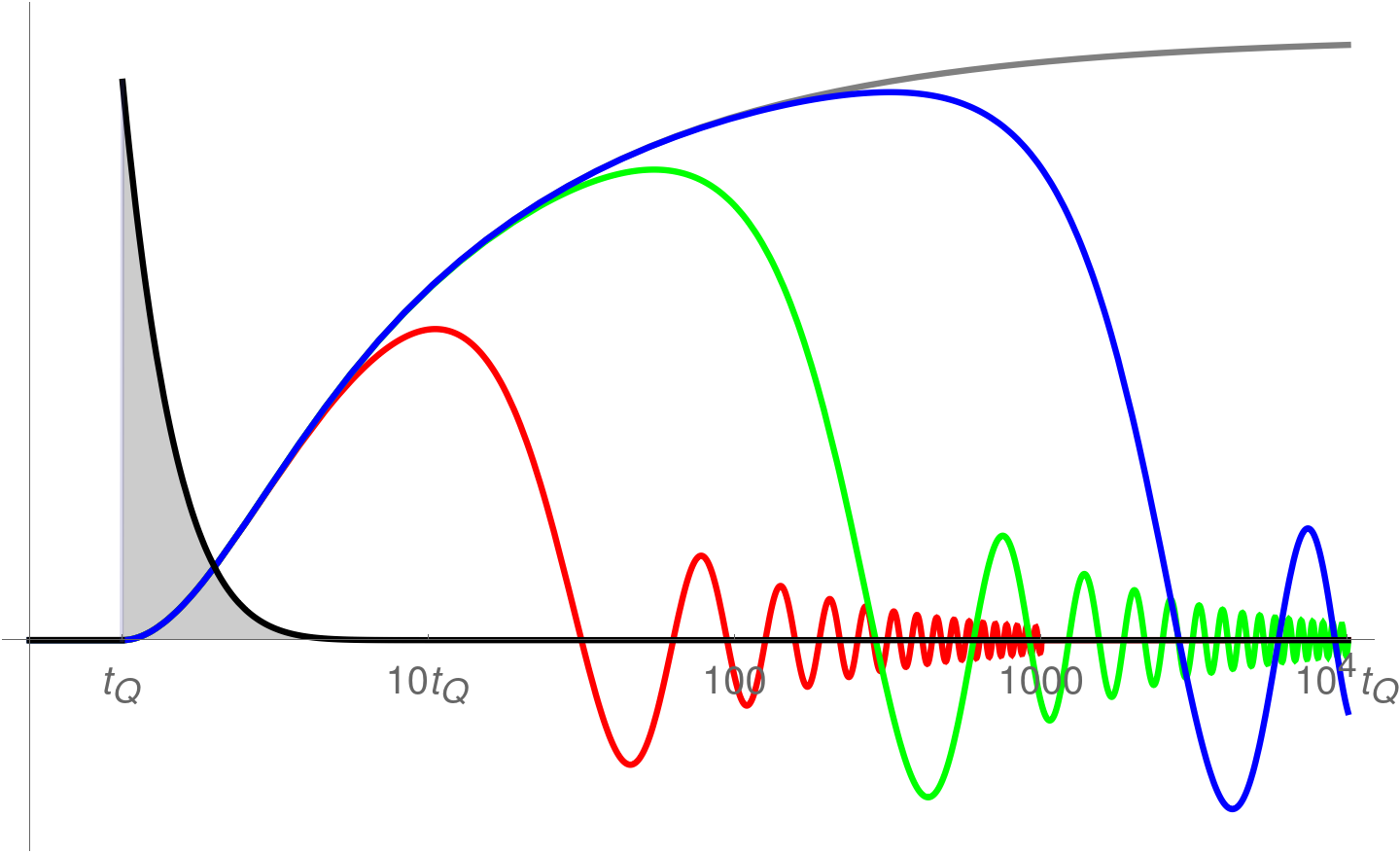}
    \caption{Black is the source term $d( a^3n_{\Delta Q})/dt/a^3$, whereas blue, green and red are $\phi(t)$ solutions for $2r_m^2=t_Q m=(0.001,0.01,0.1)$ and $r_\tau=1$. Note that for non-zero modes the evolution is obtained from the above by $m^2\to m^2+k^2$.}
    \label{fig:Kickalign}
\end{figure}

In this section we show how to calculate the $\rho_\phi$ due to Kick-alignment. Consider the cosmological evolution of the field $\phi$;
in an isotropic, homogeneous primordial universe. The dynamics of the field $\phi$ are, expanding the potential and retaining the mass term only;
\begin{align}\label{EoM}
    \frac{d^2\phi}{dt^2}+3H(t)\frac{d\phi}{dt}+(k^2+m^2)\phi&=\frac{1}{f a^3}\frac{d}{dt}(n_{Q}a^3),
\end{align}
where the RHS shows quantitatively how the breaking of  $Q$-symmetry acts as a source for $\phi$, with
\begin{align}
    n_Q \equiv J^0_Q= \sum_i Q_i\left( n_{\psi_i}-n_{\bar\psi_i}\right),
\end{align}
so for baryon number one has $n_B=n_b-n_{\bar b}$. In the following we focus on the zero mode ($k=0$) which is the last to cross the horizon and will dominate the density, see fig.~\ref{fig:Kickalign}. Given the solution to the homogeneous equation $\hat J$, the solution to the EoM with a source is
\begin{align}\label{SolEOM}
     \phi(t)=\hat J(t)\left(1+\int_{t_i}^{t} \frac{dt'}{a^3\hat J^2}\int_{t_i}^{t'} dt'' \frac{\hat J }{f}\frac{d(a^3 n_{ Q})}{dt''}\right),
\end{align}
where  the initial conditions $\phi(t_i), \dot\phi(t_i)$ are encoded in $\hat J$. During the radiation domination era (RD) this function is
\begin{align}\label{HomSL}
     \hat J(t)=C_1\frac{J_{1/4}(mt)}{(mt)^{1/4}}+C_2\frac{Y_{1/4}(mt)}{(mt)^{1/4}}\quad {\rm (RD)}
\end{align}
with $J$ and $Y$ are the Bessel functions of the first and second kind. We note and assume several things:
\begin{enumerate}
    \item Note that the second term of eq.~(\ref{SolEOM}) is invariant under re-scaling $\hat J \to \lambda \hat J$. This means that the kick acts independently of the initial amount of dark matter already present.
    \item The initial condition for $\phi$ and $\dot{\phi}$ is set by inflation/ initial misalignment. We will characterize this amount as $\phi(t_i) = \theta_0 f$. Many misalignment scenarios assume $\theta_0 = \mathcal{O}(1)$, but we will allow smaller, fine tuned values of $\theta_0 \ll 1$. This choice determines the value of $C_1$.
    \item Similarly to the misalignment scenarios the inflation tends to set $\dot\phi (t_i)\simeq 0$, which in turn implies smallness of $C_2$. We will set $C_2 = 0$.
    \item Baryogenesis happens after inflation (a fairly safe bet).
\end{enumerate}

The solution in eq.~(\ref{SolEOM}) is that of a driven damped oscillator. The presence of an external force on the oscillator is transient and lasts for the period in which the asymmetry is changing. As a result, there are several timescales in play:
The $Q$-number generation starts at $t_Q$ and lasts until $t_F$ (forcing term time). On the other hand, the field starts its oscillations by the time $t_m$ when Hubble is of the order of the mass, we implicitly define $4H(t_m) \equiv m$. The overall effect of the kick is given by the ratios of these time scales and the overall strength of the forcing term.

\begin{enumerate}
\item[A.] If the source effect ends before the first oscillation $t_F \ll t_m$ the field will get suddenly displaced from its initial value, undergo some Hubble friction and finally starts to oscillate at $t_m$ (a quick kick solution).
\item[B.] On the other hand, if the process of $Q$-number generation is slow or late $t_F \gg t_m$, the effect of the forcing term is averaged out and vanishes as can be seen in the source term in eq.~(\ref{SolEOM}) where the faster time dependence of  $\hat J\sim \cos(m t)$ makes the $t''$ integral cancel. 
\end{enumerate}
As a result, for the rest of this work we will focus on the first regime. You can see some examples of the `Kick-alignment' solutions for different choices of $t_Q/t_m$ in fig.~\ref{fig:Kickalign}. These solutions were obtained by direct integration of the EoM. The following is a discussion of the field evolution, the reader interested in the outcome at late times can skip to eq.~(\ref{Late}) for the result.

While the exact solution can be given in the form of eq.~(\ref{SolEOM}), the integral with $\hat J^{-2}$ presents (spurious) poles when the homogeneous solution starts oscillating, which make its numerical evaluation challenging. Here instead, in order to obtain the leading parametric dependence,  we will exploit the transient nature of the forcing term and
split the evolution of $\phi$ into three stages:
\begin{enumerate}
    \item[\it(i)] $\boxed{t<t_Q}$ The source term has not been 'turned on' yet so the exact solution in eq.~(\ref{SolEOM}) reduces to the homogeneous term $\hat{J}(t)$ as given in eq.~(\ref{HomSL})  satisfying the initial conditions at $t_i$.
    \item[\it (ii)] $\boxed{t_Q < t < t_F}$ While the source term is active  the field evolution is given by the initial conditions and solving the double integral in eq. \ref{SolEOM}, with $t_i = t_Q$.
    \item[\it (iii)] $\boxed{t>t_F}$ After the forcing term vanishes the evolution of $\phi(t)$ is again given by the homogeneous version of the EoM. The solution looks like eq. \ref{HomSL} with different $C_1$ and $C_2$ as compared to the solution in (i). The coefficients can be obtained by matching with the results of the previous stage at $t = t_F$. The amplitude of this solution determines the relic density of $\phi$.
\end{enumerate}
In general, we can think about the above calculation as a scattering problem: a free wave function enters a region with non-trivial potential and the final solution is again a linear combination of free wave functions with a different amplitude and a nonzero phase-shift.

As we mentioned, given our initial conditions before $Q$-number generation the field can be described as:
\begin{align}\label{Phiti}
 (i)&&  \phi(t<t_Q) &= C_1\frac{J_{1/4}(mt)}{(mt)^{1/4}}\equiv \hat J_i(t),
\end{align}
where given our assumption of $t_F\ll t_m$ and the fact that by definition $t_Q<t_F$ one has that the homogeneous solution in this regime is a constant. In other terms, the field is Hubble stuck at early times as is familiar from the misalignment mechanism. Therefore
\begin{align}
    (i)&&  \phi(t<t_Q) &=\phi(t_i).
\end{align}
During the regime (ii) the exact solution should be used and the field reads
\begin{align}\label{SolEOM2}
(ii)& \qquad (t_Q<t<t_F)\\ \nonumber
  &   \phi(t)=\hat J_i(t)\left(1+\int_{t_Q}^{t} \frac{dt'}{a^3 \hat J_i^2(t')}\int_{t_Q}^{t'} dt'' \frac{ \hat J_i(t'') }{f}\frac{d(a^3 n_{Q})}{dt''}\right)
\end{align}
this simplifies given that our assumption $t_F\ll t_m$ means $\hat J_i$ is still to a good approximation constant given by initial conditions as in eq.~(\ref{Phiti}), and so
%\begin{align}
%    r_m^2 \equiv \frac{t_F}{t_m} \ll 1
%\end{align}
\begin{align}\label{eq:stage2}
    (ii)&&\phi(t) &= \phi(t_i) + \frac{1}{f}\int_{t_Q}^{t}  n_{Q}(t)dt +\mathcal{O} \left(\frac{t_F}{t_m}\right) \\
 &&\dot{\phi}(t) &= \frac{n_{Q}(t)}{f} +\mathcal{O}\left(\frac{t_F}{t_m}\right).  
 \label{eq:stage2dp}
\end{align}
This means in particular that the source has generated velocity in $\phi$ and there is a contribution to $Y_{1/4}$ in the matching, which to first order in each coefficient reads:
\begin{align}\label{matchF}
    (iii)&& \phi(t)&= \bar C_1 \frac{J_{1/4}(mt)}{(mt)^{1/4}}+\bar C_2\frac{Y_{1/4}(mt)}{(mt)^{1/4}}\\
    && \bar C_1&=2^{1/4}\Gamma(5/4)\left(\phi(t_F)+2t_F\dot\phi(t_F)\right)\\
    &&\bar C_2&=\frac{2^{5/4}\pi}{\Gamma(1/4)} t_F\dot\phi(t_F) \left(\frac{t_F}{t_m}\right)^{1/2}
\end{align}
where $\phi(t_F)$, $\dot\phi(t_F)$ are expressions in eq.~(\ref{eq:stage2},\ref{eq:stage2dp}) evaluated at $t_F$ and this function describes the remainder of the evolution of the field, in particular at late times after horizon crossing one has
\begin{align}\label{Latte}
    \phi(t\gg t_m)= \sqrt{\frac{\bar C_1^2+\bar C_2^2}{\sqrt{2}\pi}}\left(\frac{a(t_m)}{a(t)}\right)^{3/2}\cos(mt+\omega).
\end{align}
Hence our approximation $t_F/t_m\ll1$ means that at late times (but not intermediate ones!) the amplitude is given by $\bar C_1$.

An important fact to stress is that, provided $t_F\ll t_m$ the value of $t_F$ at which we do the matching should not matter. This is evidenced by the fact that the dependence on $t_F$ on $\bar C_1$ can be rewritten as
\begin{align}\nonumber
    \phi(t_F)+2t_F\dot\phi(t_F)&=\phi(t_i)+\int_{t_Q}^{t_F}\frac{n_{Q}(t)}{f}dt+\int_{t_F}^{\infty}\frac{n_{Q}(t)}{f}dt\\
    &=\phi(t_i)+\int_{t_Q}^{\infty}\frac{n_{Q}(t)}{f}dt,
\end{align}
where we used the fact that by $t_F$ the time dependence  of $n_Q$ is $\propto a^{-3}$. In view of the above and substitution back in eq.~(\ref{Latte}) via eq.~(\ref{matchF}) one can see that the effect at late times can be encoded in a shift on the initial value of the field by an amount $\delta\phi$ given by the integral of $Q$ abundance. 
While this much one could have guessed from the exact solution eq.~(\ref{SolEOM}) assuming in the inner integral $a^3n_Q$ has a step-function form and the outer integral converges due to the $a^{-3}$ factor, this derivation as presented makes it explicit and more general.

The net effect of a shift in the field is the reason we call this mechanism the `Kick-alignment', and the fact that this shift is calculable and related to observed physics makes it more predictive than its well studied counterpart, the misalignment. Nonetheless, for the shift to dominate one has to assume that the initial value $\phi(t_i)$ is much smaller, $\phi(t_i)\ll \delta \phi$ which in certain cases might imply fine-tuning (unless a dynamic mechanism is put in play to set $\phi(t_i)=0$ but we will not theorize about this here). In the following, we assume  $\phi(t_i)\ll \delta \phi$ holds since in the opposite case, the dominant initial value leads to conventional and exhaustively studied misalignment. 

In order to make the potential fine-tuning explicit we cast the integral for $\delta\phi$ in terms of observed quantities for clearer display of its order of magnitude 
\begin{align}\label{deltaphi}
   \delta\phi \equiv& \frac{1}{f}\int  n_{Q}(t)dt  +\mathcal O\left(\frac{t_{F}}{t_m}\right) 
   %\frac{1}{f} \int \frac{dT}{HT}\left(\frac{n_{ Q}}{s}\right)s
   \\\nonumber
   =&  \sqrt{\frac{\pi}{45}}\frac{\ate{Q} \sqrt{g_\star(T_Q)}M_{\textrm{pl}}T_Q}{f}\int \frac{dT}{T_Q} \sqrt{\frac{g_\star(T)}{g_\star(T_Q)}} \frac{ \ate{Q}(T)}{\ate{Q}(T_0)}\\\label{DisPl}
   &+\mathcal O\left(\frac{t_{F}}{t_m}\right) 
   \\
   \equiv& \sqrt{\frac{\pi}{45}}\frac{\ate{Q}\sqrt{g_\star(T_Q)} M_{\textrm{pl}}T_Q}{f} \,I(r_m,r_\tau),
\end{align}
where we define $\ate{Q} \equiv n_Q/s$ and the last line defines  $I(r_m, r_\tau)$ which includes sub-leading corrections and is a function of variables
\begin{align}\label{QK}
    r_m^2&=\frac{t_Q}{t_m} & r_\tau^2&=\frac{t_Q}{\tau}
\end{align}
where we introduced $\tau$ as the typical scale of $Q$ production, (e.g. the lifetime of a heavy particle decaying asymmetrically) and one has that $t_F-t_Q=$*several*$\tau$ so that the process as ended by $t_F$ (e.g. the effect has decreased by *several* e-folds for the heavy particle decay case). The justification for this extra notation is that as underlined late time results should not depend $t_F$ but they will depend on $\tau$. In addition the first order correction on $r_m$ can be captured selecting the limits of integration as shown in appendix~\ref{sec:matching}.

This integral is order 1 for $\tau\sim t_Q$ so for order of magnitude estimates one can take the shift to be, in terms of the angle variable $\theta=\phi/f$:
\begin{align}
       \delta\theta=\frac{\delta\phi}{f}\simeq \ate{Q}\sqrt{g_\star} T_Q \frac{M_{\rm pl}}{f^2}.
\end{align}
Given the the matter anti-matter asymmetry today one therefore has and order one angle displacement for $f\sim \sqrt{T_Q/{\rm GeV}}10^5$ GeV. 
At the same time, we note that if the displacement of the field is large enough $\delta \phi \sim f$ one can not expand the potential to leading order and the non-linear EoM should be solved. This regime is reached for $f^2\sim \ate{Q} M_{pl} T_Q$; if $f$ is near $T_Q$ this implies $f\sim 10^{9}$ GeV.
%%%%%%%%%%%%%%%%%%%%%%%%%%%%%%%%%%%%%%%%%%%%%%%%%%%%%%%

Plugging in the above definitions back into the late time behaviour of the field in eq.~(\ref{Latte}) reads 
\begin{align}\label{Late}
    \phi(t)=\frac{\Gamma_{5/4}}{\sqrt{\pi}}\delta\phi\left[\frac{a(t_m)}{a(t)}\right]^{3/2}\cos(mt+\omega)
\end{align}
with $\Gamma_{5/4}\equiv\Gamma(5/4)\simeq 0.906$. The field has evolved to have an amplitude today as
\begin{align}
    \phi(t_0)= &A(t_0) \cos(mt_0+\omega)\\
    A(t_0)^2=&\frac{\Gamma^2_{5/4}}{\pi}\frac{43}{11g_\star}\left(\frac{ 64\pi^3g_\star T_0^4}{45M_{\rm pl}^2m^2}\right)^{3/4}(\delta\phi)^2\\
    =&\frac{\Gamma^2_{5/4}}{\pi} \left(\frac{180\pi}{g_{\star}(t_m)}\right)^{1/4} \frac{8s_0(\delta\phi)^2}{(m M_{\rm pl})^{3/2}} \,
\end{align}
so the energy density, $\rho=A^2m^2/2$, is
\begin{align}\nonumber
    \rho_\phi=&\frac{4\Gamma^2_{5/4}(4\pi)^{1/4} g_\star(T_Q)}{(45)^{3/4}g_\star(t_m)^{1/4}} \ate{Q} \sqrt{m M_{\rm pl}}\frac{T_Q^2}{f^2} I^2 n_Q(t_0)\\
    =&0.356  \frac{g_\star(T_Q)}{g_\star(t_m)^{1/4}} \ate{Q} \sqrt{m M_{\rm pl}}\frac{T_Q^2}{f^2} I^2 n_Q(t_0)\label{rhoCh}
\end{align}
%with 
%\begin{align}
    %I(r_\tau,r_m)%=\frac{C}{2}
    %=\int_{r_mT_Q}^{T_Q} \frac{dT}{T_Q} %\sqrt{\frac{g_\star (T)}{g_\star (T_Q)} } %\frac{\Delta N_Q(T)}{\Delta N_Q (T_0)}
%\end{align}
%\JS{Let us include the amplitude correction into the integral.}
%where a quick or step-wise $Q$ generation returns $I$ of order one, the precise value given by tracing the redistribution of $Q$ number from generation till today.

Assuming $T_Q\sim f$ and $I\sim 1$, the right abundance for $\phi$ to make up the dark matter can obtained when the geometric mean mass of $m$ and $M_{\rm pl}$ times the asymmetry $\ate{Q}$ falls around the mass of baryons themselves. Which is to say the right relic abundance is obtained for a value 5.4 ($\Omega_{dm}\sim 5.4\Omega_b$ \cite{Akrami:2018vks}) of the ratio: 
\begin{align}\nonumber
    \frac{\rho_{\phi}}{m_bn_b}= &0.356 \frac{g_{\star}(T_Q)}{g_\star(t_m)^{1/4}}\frac{\ate{Q} \sqrt{m M_{\rm pl}}}{m_b} \frac{T_Q^2}{f^2}\frac{n_{Q}(t_0)}{n_b(t_0)}I^2\\
    &=\frac{0.12g_{\star}(T_Q)}{g_\star(t_m)^{1/4}} \sqrt{\frac{m}{\rm GeV}} \frac{T_Q^2}{f^2}\frac{\ate{Q}n_{Q}(t_0)}{\ate{B} n_b(t_0)}I^2\label{RelicA}
\end{align}
The fact that the mass of dark matter falls around the electro-weak scale merits some mention even if we will not speculate on why this could be the case here. What does follow unambiguously from the expression above is that there is a lower limit on the mass $m$ if this is to be the dark matter, which we take here to be given by the theory constraint $T_Q< f$ (i.e. spontaneous symmetry breaking to occur before $Q$ generation).

The `Kick-alignment' production of $\phi$ is then analytically calculable within the approximations here considered yet it occurs in more general circumstances.
This work however aims at showing the feasibility of the mechanism rather than exhausting its possibilities so let us proceed
to an analysis with explicit realizations of $B$ and $L$ generation. This will give physical input on $T_Q$ and $I$ while providing useful nontrivial consistency conditions.

\subsection{$Q$-number generation mechanisms}
Here we consider the possibilities of $Q$ violation from a heavy particle decay or alternatively through Sphaleron transmission from another sector.\\[2mm]
%%%%%%%%%%%%%%%%%%%%%%%%%%%%%%%%%%%%%%%%%%%%%%%%%%%5
(I)$Q$  {\bf generation in heavy particle decay }\\\noindent
Let some heavy particle $X$ decay out of equilibrium and in different proportion to particles and antiparticles.
The departure of equilibrium sets $t_Q$ and we parametrise the asymmetry generation as ($N_{X}\sim e^{-\Gamma_X t}$) $ \ate{Q}\propto (1-e^{-\Gamma_X(t-t_Q)})$ with the identification $\tau=\Gamma^{-1}_X$ in eq.~(\ref{QK}). The ratio of asymmetry reads then:
\begin{align}
    \frac{ \ate{Q}(t)}{ \ate{Q}(t_0)}=\beta(1-e^{-\Gamma_X(t-t_Q)})
\end{align}
where $\beta$ is an order one number and accounts for redistribution factors of $Q$ symmetry from generation to present time typically given by Sphaleron processes (e.g. for $Q=L$ and Leptogenesis $\beta= 1/(1-12/37)$). 
The integral $I$ in terms of temperature thus reads (assuming RD, $\dot T=-HT$)
\begin{align}
    I(r_\tau,r_m)=&\beta\int \frac{dT}{T_Q} \left(1-e^{-r_\tau^2((T_Q/T)^2-1)} \right),\label{IX}
    %\\ =& \beta\int^1_{0.9r_m} dr \left(1-e^{-r_\tau^2(r^{-2}-1)}\right)
\end{align}
where the time and temperature for the start of $Q$ generation is determined by the particle $X$ falling out of equilibrium, $T_Q=M_X/x_f$ with $x_f$ given by freeze-out. 
The integral $I$ can be rewritten in terms of a Gaussian integral and the exact expression is given in  appendix~\ref{App:Int}.
One can estimate $x_f$ as the temperature when the inverse decay process becomes inefficient compared to Hubble, that is 
\begin{align}
\Gamma_X e^{-x_f}&\simeq H(x_f) & K\equiv&\frac{\Gamma_X}{H(M_X)}
\end{align}
One has then an early departure from equilibrium $x_f\sim 1$ occurs for weak coupling $K\ll 1$ or a slightly later one $x_f\sim\log(K)$\cite{Davidson:2008bu}  occurs for strong coupling $K\gg1$. The strong coupling larger rate however makes $Q$ number generation resemble a step-function whereas the slower growth for $K\ll 1$  results in a smaller value for the integral. This translates into the two asymptotic behaviors
\begin{align}\nonumber
    K&\gg 1 & r_\tau^2&\simeq \frac{K\log^2(K)}{2} & T_Q&\simeq \frac{M_X}{\log(K)} & \frac{I}{\beta}&\simeq1\\\nonumber
    K&\ll 1& r_\tau^2&\simeq \frac{K}{2} & T_Q&\simeq M_X &\frac{I}{\beta}&\simeq\sqrt{\pi} r_\tau 
\end{align}
where we took $r_m\ll1$.
Both limits $K\gg1$ and $K\ll 1$ will be explored in the phenomenology in section~\ref{RealZ}.\\[2mm]
%%%%%%%%%%%%%%%%%%%%%%%%%%%%%%%%%%%%%%%%%%%%%%%%%%%
 (II)$ Q$ {\bf generation by Sphaleron transfer}\\\noindent 
It could be the case instead that e.g. a $L$ violating decay produces an initial asymmetry while  $Q=B$. It would only be when Sphaleron processes transfer the asymmetry to $B$ number that $\phi$ starts feeling a source. Following this example the evolution in $ \ate{B}$ reads \cite{KHLEBNIKOV1985}
\begin{align}
    \frac{d(\ate{B})}{ dt}&=-\kappa \alpha_w^4 c_1T\left( \ate{B}- c_2 \ate{B-L}\right)\\
    c_1=&N_f^2\frac{3}{4}\frac{22N_f+13}{N_f(5N_f+3)}\quad c_2=\frac{8N_f+4}{22N_f+13}
\end{align}
where we assume $T>T_{EWPT}$ with the generations $N_f=3$ and the initial $\ate{B-L}$ is taken as an input.
The time scale for this transfer of asymmetry is given by the sphaleron rate and with the identification \begin{align}
    \tau^{-1}=\kappa c_1\alpha_{\rm ew}^4 T_Q
\end{align} one has a $ \ate{B}$ dependence with temperature for the solution
\begin{align}
    \ate{ B}%&\simeq\frac{c_2}{c_1} N_{\Delta B-L}(1-e^{\alpha_w T(t_Q)a(t_Q)(\eta-\ate{Q})})\\
    &=c_2 \ate{B-L}\left(1-e^{-2r_\tau^2(T_Q/T-1) }\right)%\\
    %K_{\rm sp}&\equiv\frac{\kappa\alpha^4_{\rm ew} T_i}{H(T_i)}
\end{align}
taking now $T_Q$ as the temperature when the initial $B-L$ asymmetry is generated. Hence our integral
\begin{align}\label{IS}
    I=\int \frac{dT}{T_Q} (1-e^{-2r_\tau^2((T_Q/T)-1)})
\end{align}
where the ratio $r_\tau^2$ now relates to whether Sphalerons are in equilibrium or not by $T_Q$; in particular the two limits for $r_\tau$ return 
\begin{align}
r_\tau^2=\frac{\Gamma_{\rm sp}(T_Q)}{2H(T_Q)}&\gg1 & I&=1\\
r_\tau^{2}=\frac{\Gamma_{\rm sp}(T_Q)}{2H(T_Q)}&\ll1 & I&=\left(\log\left(\frac{1}{2r_\tau^2}\right)-\gamma_E\right)2r_\tau^2
\end{align}
with $\gamma_E$ the Euler-Mascheroni constant. The full solution without neglecting $r_m$ is given in terms of the incomplete Gamma function in appendix~\ref{App:Int}.

The discussion of sec.~\ref{Sec:TH} suggests that if Sphalerons can source the asymmetry, thermal production might overcome Kick-alignment and the subsequent phenomenological analysis focuses on case (I) for simplicity.

It is clear, nonetheless, that if one has a Goldstone boson for a $Q$ symmetry
there will be $Q$-violating-sourced production of the field. Whether this is a sufficiently strong generation mechanism is then a quantitative question.

%%%%%%%%%%%%%%%%%%%%%%%%%%%%%%%%%%%%%%%%%%%\subsection
\subsection{Back reaction on $\ate{Q}$}
A relevant point to address is the feedback into the asymmetry $ \ate{Q}$ of the field $\phi$.
The effect of a non-vanishing time derivative of $\phi$ creates a non-zero chemical potential for the fermions that make up $J_Q$ as can be seen in Dirac's equation:
\begin{align}
    \left[ iD\!\!\!\!\slash-(m+\gamma^0Q_\psi \dot \phi/f)\right]\psi=0.
\end{align}
This is the principle behind Spontaneous Baryogenesis \cite{Cohen:1987vi,Cohen:1988kt,Cohen:1993nk}.
Nevertheless this contribution by itself is not observable (as again the field rotation $\psi\to e^{-iQ\phi/f}\psi$ from~(\ref{Rotphi}) exemplifies), the other necessary physics for an effect is an explicit $Q$-violation. Through this $Q$-violating effect, the potential $\mu_\phi=\dot \phi/f$ will contribute to an particle anti-particle imbalance. For an estimate of the effect we compare this generated chemical potential with the potential that produced the asymmetry in the first place,
\begin{align}
\frac{\mu_0}{T}\simeq &\frac{n_{Q}}{n_q} &
    \frac{\mu_\phi}{T}&\simeq \frac{\dot \phi}{fT}\simeq \frac{n_{ Q}}{f^2 T} 
\end{align}
where we used $n_Q=n_q-n_{\bar q}$ and so $\mu_0 > \mu_\phi$ provided $f>T$ and hence the back-reaction from the presence of $\phi$ is negligible by default because this is the condition we imposed already so that the theory contains a Goldstone boson $\phi$ by the time the Universe reaches temperature $T_Q$. 

Nevertheless, we believe that the scenario in which the feedback from the fermions is in some sense in equilibrium with the effect we describe here could be of theoretical interest and leave this possibility for future study.
%%%%%%%%%%%%%%%%%%%%%%%%%%%%%%%%%%%
\subsection{ Contrast with other production mechanisms}
Below we compare the parametric dependence of various DM production mechanisms: freeze-out (FO), freeze-in (FI), the misalignment (MA) and the Kick-alignment (KA). Taking the yield $Y$ as $\rho/(m T^3)$ one has the estimates for each mechanism as
\begin{align}\nonumber\textrm{FO:} & \quad Y\propto\frac{1}{\langle\sigma v\rangle M_{\rm pl}m}&\textrm{FI:} &\quad Y\propto\alpha^2 \frac{M_{\rm pl}}{m}\\ \nonumber\textrm{MA:} &\quad Y\propto\frac{\theta_0^2f^2}{M_{\rm pl}\sqrt{M_{\rm pl} m}}  &\textrm{KA:}&\quad Y\propto\ate{B}^2\frac{T_Q^2}{f^2}\sqrt{\frac{M_{\rm{pl}}}{m}}\end{align}
The arrangement is meant to be meaningful: we have thermal production on the first row and athermal in the second while the second column has a yield  proportional to the coupling, the first column's yield is inversely proportional to the coupling. In this sense it is clear that the Kick-alignment completes the square.

Notice that Freeze-in operates in situations where the initial population can be neglected, as opposed to the Freeze-out, and hence can works with small couplings. The same way the Kick-alignment works best when the initial field value is small, exactly opposite to the case of misalignment. In that sense the Kick-alignment is to misalignment as Freeze-in is to Freeze-out. 

%%%%%%%%%%%%%%%%%%%%%%%%%%%%%%%%%%%

\subsection{Thermal production}\label{Sec:TH}
The coupling of $\phi$ can be reshuffled to baryon or lepton violating interactions via a $\phi$-dependent symmetry transformation, as shown in eqs.~(\ref{Rotphi},\ref{eq:psiQ}). Since the SM has no relevant B or L violating operators, the $Q$-violating interactions have to be non-renormalizable at energies around the Standard Model. These type of interactions induce thermal production dominated by the UV scale. This introduces model dependence yet it also means that the field is easy to produce at early times whereas at low energies its couplings are very suppressed.

There is one such production channel with $B+L$ violation in the SM in Sphaleron processes so let us being by estimating this.

\begin{figure}[ht]
    \centering
\begin{tikzpicture}
\filldraw (0,0) circle (4pt);
\draw [thick] (-1,0) -- (0,0);
\draw [>-,thick] (-0.5,0) -- (0,0);
\draw [thick] (-0.85,1/2) -- (0,0);
\draw [>-,thick] (-0.85/2,1/4) -- (0,0);
\draw [thick] (-0.85,-1/2) -- (0,0);
\draw [>-,thick] (-0.85/2,-1/4) -- (0,0);
\draw [thick] (-1/2,0.85) -- (0,0);
\draw [>-,thick] (-1/4,0.85/2) -- (0,0);
\draw [thick] (-1/2,-0.85) -- (0,0);
\draw [>-,thick] (-1/4,-0.85/2) -- (0,0);
\draw [thick] (0,-1) -- (0,0);
\draw [>-,thick] (0,-1/2) -- (0,0);
\draw [thick] (0,1) -- (0,0);
\draw [>-,thick] (0,1/2) -- (0,0);
\draw [thick] (1/2,0.85) -- (0,0);
\draw [>-,thick] (1/4,0.85/2) -- (0,0);
\draw [thick] (1/2,-0.85) -- (0,0);
\draw [>-,thick] (1/4,-0.85/2) -- (0,0);
\draw [thick] (0.85,1/2) -- (0,0);
\draw [>-,thick] (0.85/2,1/4) -- (0,0);
\draw [thick] (0.85,-1/2) -- (0,0);
\draw [>-,thick] (0.85/2,-1/4) -- (0,0);
\draw [thick] (1,0) -- (0,0);
\draw [>-,thick] (1/2,0) -- (0,0);
\draw [dashed] (0,0) -- (0.7,-0.7) node [anchor=north west] {$\phi/f$};
\draw [thick] (2,0.85) -- (2.5,0);
\draw [>-,thick] (2.25,0.85/2) -- (2.5,0);
\draw [thick,dashed] (2.5,0) --(3,-0.85);
\draw [-<,thick,dashed] (2.5,0) --(2.75,-0.85/2);
\draw [thick] (2,-0.85) --(2.5,0);
\draw [>-,thick] (2.25,-0.85/2) --(2.5,0);
\draw [thick,dashed] (2.5,0)--(3,0.85);
\draw [-<,thick,dashed] (2.5,0)--(2.75,0.85/2);
\filldraw (2.5,0) circle (2pt);
\draw[dashed] (2.5,0) -- (3.5,0) node [anchor=north] {$\phi/f$} ;
\draw [thick] (4.5,0.85) -- (5,0);
\draw [>-<,thick] (4.75,0.85/2) -- (5.25,-0.85/2);
\draw [thick] (5,0) --(5.5,-0.85);
\draw [thick] (4.5,-0.85) --(5,0);
\draw [>-<,thick] (4.75,-0.85/2) --(5.25,0.85/2);
\draw [thick] (5,0)--(5.5,0.85);
\filldraw (5,0) circle (2pt);
\draw[dashed] (5,0) -- (6,0) node [anchor=north] {$\phi/f$} ;
\end{tikzpicture}
    \caption{$\phi$ coupling for thermal production via Sphalerons, dimension 5 ($L\!\!\!\slash$) and 6 ($B\!\!\!\slash$) operators.}
    \label{fig:FYNR}
\end{figure}
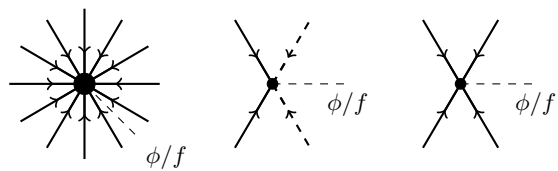
The temperature dependence of the rate for Sphaleron distinguishes two regimes. After EWSB the rate is dominated by an exponential of minus the Sphaleron energy over the temperature whereas at higher temperatures the tunneling suppression is lifted and the rate per unit volume scales with
\begin{align}
\frac{\Gamma_{\rm sp}}{V}& \sim \kappa (\alpha_w T)^4 & T&\gg T_{EWPT}
\end{align}
if we rotate the fermions as in eq.~(\ref{Rotphi},\ref{eq:psiQ}) to cancel the coupling $\partial_\mu\phi J^\mu$ the field will appear instead on the 't Hooft vertex as a phase factor, which when expanded, yields a linear coupling of $\phi/f$ as sketched in fig.~\ref{fig:FYNR}. Assuming then the reheat temperature is greater than a few TeV, $f\gg T$ we expect the thermal production to be dominated by the Sphaleron processes so
\begin{align}
a^3\frac{d a^3 n_\phi}{dt} \simeq \kappa (\alpha_w T)^4 \left(\frac{T}{4\pi f}\right)^2    
\end{align}
assuming the field was initially out of equilibrium and that it never reaches it, we are in the regime of UV freeze-in (otherwise we have to include the inverse processes as well). In terms of temperature this thermal production reads ($Y_\phi=n_\phi/T^3$):
\begin{align}
    \frac{d Y_\phi}{dT}\simeq-\sqrt{\frac{45}{4\pi^3 g_\star}}\frac{\kappa \alpha_w^4 M_{\rm pl}  }{(4\pi f)^2}
\end{align}
which confirms the yield is dominated by the UV (high $T$). This thermal contribution to energy density is then:
\begin{align}
\rho_\phi &  \simeq m_\phi\frac{43}{11g_\star} \sqrt{\frac{45}{4\pi^3 g_\star}}T_0^3 \frac{\kappa \alpha_w^4 M_{\rm pl} T_{rh}}{(4\pi f)^2} %\\
   % \Omega & \sim \left(\frac{m_\phi}{\rm GeV}\right) \left(\frac{10^{3} \kappa \alpha_w^4 M_{\rm pl}T_{rh} }{f^2}\right)
\end{align}
Comparison with Kick-alignment production shows how the different scaling with $m$ implies that for heavy $m$ this thermal mechanism dominates. 
This poses a strong constraint on successful Kick-alignment. This can be avoided taking our symmetry as $Q=B-L$ or any other combination to be preserved in the SM.

Other thermal channels of $\phi$ production will equally be dominated by high temperature but the specifics follow the mechanism by which the baryon asymmetry of the Universe was generated. To estimate this production let us consider the leading operators violating L and B are:
\begin{align}
C_5\mathcal O_5+C_6\mathcal O_6=\frac{y_N^2}{2M_N}(\ell H)^2 +\frac{y_X^2}{M_X^2}q q q \ell.
\end{align}
These will produce the vertices for $\phi$ interaction as shown in fig~\ref{fig:FYNR}.
This simplistic picture still allows for nontrivial constraints on Kick-alignment production once the couplings and masses in these operators are identified with the $Q$ generation mechanism.

The thermal production of $\phi$ from these non-renormalizable operators reads c.f.~\cite{Hall:2009bx};
\begin{align}
\frac{d Y_\phi}{dT}=-\sqrt{\frac{45}{4\pi^3 g_\star}}\frac{M_{pl}}{16\pi^5(4\pi f)^2}\left( C_{d}T^{d-4}\right)^2
\end{align}
In particular with the definition of the operators above
\begin{align}
    Y_\phi&\simeq \sqrt{\frac{45}{4\pi^3 g_\star}}\frac{y_N^4M_{\rm pl}M_N}{3\times 64\pi^5(4\pi f)^2}\sin^2\alpha%=\sqrt{\frac{90}{8\pi^3 g_*}}\frac{M_{pl}m_\nu^2 M_N^3}{192 v^4\pi^5(4\pi f)^2}
    \\
    Y_\phi&\simeq \sqrt{\frac{45}{4\pi^3 g_\star}}\frac{y_X^4M_{\rm pl}M_X}{5\times 16\pi^5(4\pi f)^2}\sin^2 (\alpha- \pi/4)
\end{align}
were $v=174$~GeV. If one assumes $\mathcal O_5$ is also the sole contribution to  light neutrino masses the relation $m_\nu = y_N^2 v^2/M_N$  holds and can be used to further constrain the parameter space.
%%%%%%%%%%%%%%%%%%%%%%%%%%%%%%%%%%%%%%%%%%%%%%%%%%%
\section{Realization}\label{RealZ}
In order to evaluate the feasibility of the Kick-alignment production mechanism proposed in the previous section we will explore particular cases of leptogenesis and baryogenesis. In particular we will consider the two scenarios with two different $Q$ symmetries 
\begin{align}\nonumber
\textrm{A.}\,\,&\textrm{GUT-like Baryogenesis} &\textrm{B.}\,\,&\textrm{Leptogenesis}\\ &Q=B-L &  &Q=B-3L_e\nonumber
\end{align}
both of which are exactly preserved in the Standard Model in in particular there will be no Sphaleron thermal contribution.

It should be remarked, nonetheless, that the matter asymmetry production mechanisms are taken as general archetypes rather than specific models. They serve as reference points and inspiration for more detailed future studies.
\subsection{GUT-like Baryogenesis}
In this case the baryon violating operator is the standard $\mathcal{O}_6 = qqq\ell$, where we take the strength of the operator to be $y^2_X\sim M_X/M_{\rm pl}$  which implies small $K= \Gamma_X/H_X\sim 1/4\pi\sqrt{g_\star}$. The bound on baryon number violation processes from proton decay $y_X^2/M_X^2\leq 1/(10^{16}$GeV$)^2$ can be combined combined to give $M_X\geq 10^{13}$~GeV. This very stringent bound from proton decay implies both that the lifetime of $\phi$ is well beyond the age of the universe and the thermal production through the operator $\mathcal O_6$ is suppressed up to high masses $m$.

As a result, the allowed range of parameters for which $\phi$ is produced through Kick-alignment is large and easily exceeds the thermal generation. The main constraints are: a) upper bound on $f$ from non-perturbative gravitational contributions to the mass $m$, b) the combination of $y_X^2\sim M_X/M_{\rm pl}$, c) proton lifetime bound for a lower bound on $f$, and d) the overtaking of thermal contributions for large $m$.

\begin{figure}
    \centering
    \includegraphics[width=.45\textwidth]{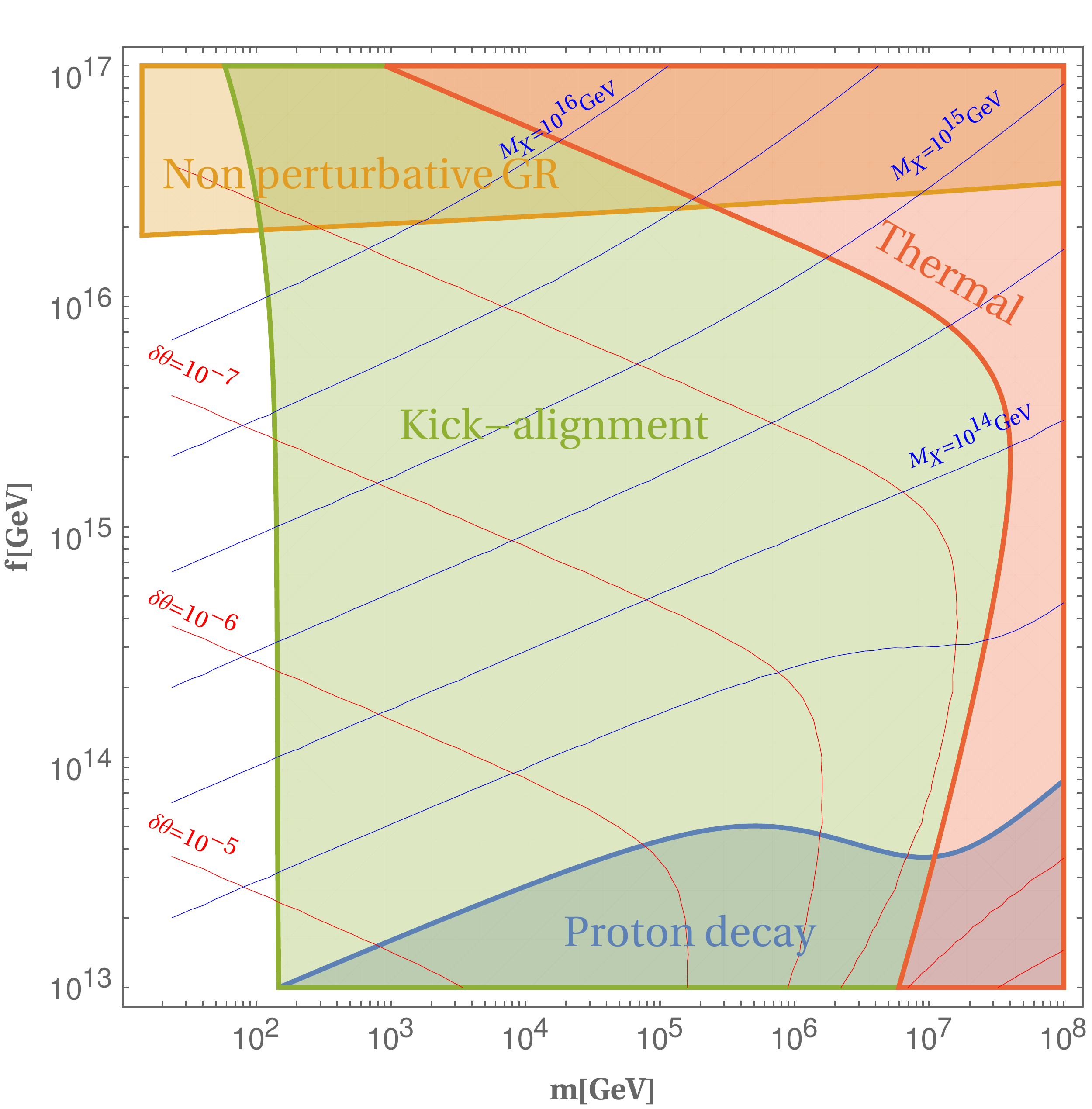}
    \caption{GUT-like asymmetry generation with $y_X^2\sim M_X/ M_{\rm pl}$ and $Q=B-L$. See text and labels for details. }
    \label{fig:GUT}
\end{figure}

For $f$ close to the Planck scale the gravitational non-perturbative contributions to the mass of $\phi$ as taken from \cite{Alonso:2017avz} are sizable and imply a $f-m$ correlation as show by the yellow line on fig~\ref{fig:GUT}. This is to be taken as a conservative upper limit in the absence of a radial mode, its inclusion generically makes the bound stronger. Alternatively, a point lying on this line can be taken as a dark matter candidate whose mass is purely gravitational. It is however important to emphasize that this is a 'theory bound' as opposed to an experimental bound.

The lower bound on $f$ follows from fitting requiring correct relic abundance (a sum of Kick-alignment and thermal production) together with proton lifetime constraints. Where the Kick-alignment dominates over thermal production the relation $T_Q\propto M_X$ implies that for a given mass $m$,  $f$ cannot be lower than $\sim T_Q\sqrt{m/{\rm GeV}}\simeq M_X\sqrt{m/{\rm GeV}}$ or the abundance of $\phi$ would be larger than observed. This bound is shown by the blue-shaded region in fig~\ref{fig:GUT}. In the red region thermal production overcomes Kick-alignment production. Finally, the low mass limit comes from requiring a good EFT expansion $f>T_Q$.
Lines of constant field shift and heavy particle mass $M_X$ are displayed in red and blue respectively and extend to the left as far as $\pi f/2>T_Q$. Note, that this internal consistency condition is, as parametrised in our plot, somewhat sensitive to the exact choice of the boundary because  $\rho_\phi = {\rm const}  \propto m^{1/2}f^{-2}$, and so requiring $\pi f>T_Q$ instead, changes the `allowed' mass range by a factor of 16. However, this is to be expected, because at this point additional degrees of freedom are relevant and the calculation is unreliable without a full model.
%This modified condition, while still yielding a acceptable EFT expansion, shows how sensitive the lower mass limit is to changes in $f$ as follows from the scaling $\rho_\phi\sim m^{1/2}f^{-2}$.
\subsection{Leptogenesis}

For Leptogenesis the field $\phi$ resembles the Majoron as DM \cite{Garcia-Cely:2017oco,Rothstein:1992rh,Berezinsky:1993fm}, albeit with a different production mechanism given that typically conventional freeze-in is considered \cite{Frigerio:2011in}.

We assume that the heavy neutrinos that give rise to the asymmetry also produce the active neutrino masses for a more constrained scenario. 
The coupling of the field  $\phi$ through the lepton-number violation operator $\mathcal O_5$ is then given by neutrino masses and the effective scale is now much lower than for Baryon number violation through $\mathcal O_6$. This implies 
stronger constraints from thermal production and from the cosmic scale stability of $\phi$. One finds conversely that now $K\sim y_N^2M_{\rm pl}/M_N\sim M_{\rm pl} m_\nu/v^2\gg 1$ which yields an $I \approx 1$.  

The lifetime and thermal production of $\phi$ are related to neutrino masses by the effective neutrino mass $m_\nu^{\rm eff}$ that the lepton part of our symmetry aligns with. This is $(m_\nu^{\rm eff})^2=\sum m_{\nu_i}^2$ for flavour blind lepton number. This possibility seems in tension with the constraints so instead we select the symmetry $3L_e$ which via a mild suppression in $m_\nu^{\rm eff}$ allows successful Kick-alignment production. However, there is not much room for evading constraints, since in the decoupling limit the Kick-alignment production vanishes. This in turn, leads to testable consequences and provides a well defined ballpark for $f$ and $m$. For the present symmetry in the lepton sector  $B-3L_e$, one has
\begin{align}
    \left(m_\nu^{\rm{eff}}\right)^2= \frac12\left|3U_{ei}^2m_i\right|^2+\frac12\sum_\beta \left|3U_{ei}U_{\beta i}m_{\nu_i}\right|^2
\end{align}
We adopt normal hierarchy and a massless first generation neutrino. This implies the contribution from the heaviest neutrino $m_{\nu_3}\sim \Delta m_{\rm atm}$ suppressed by $\theta_{13}$. This suppression can be further aided by a selection of the Majorana phase $\alpha$ in
\begin{align}
    m_\nu=\textrm{Diag}(0\,,\, m_{\nu_2}e^{i\alpha}\,,\,m_{\nu_3})
\end{align}
 of $\alpha=-0.6$. However, this does not lead to significant fine-tuning. All in all, this means $m_\nu^{\rm{eff}}\simeq 0.017$~eV taking central values for neutrino mass parameters from~\cite{Zyla:2020zbs}. For reference this gives a prediction for neutrino-less beta decay $m_{\beta\beta}=3$~meV.
 
\begin{figure}
    \includegraphics[width=.45\textwidth]{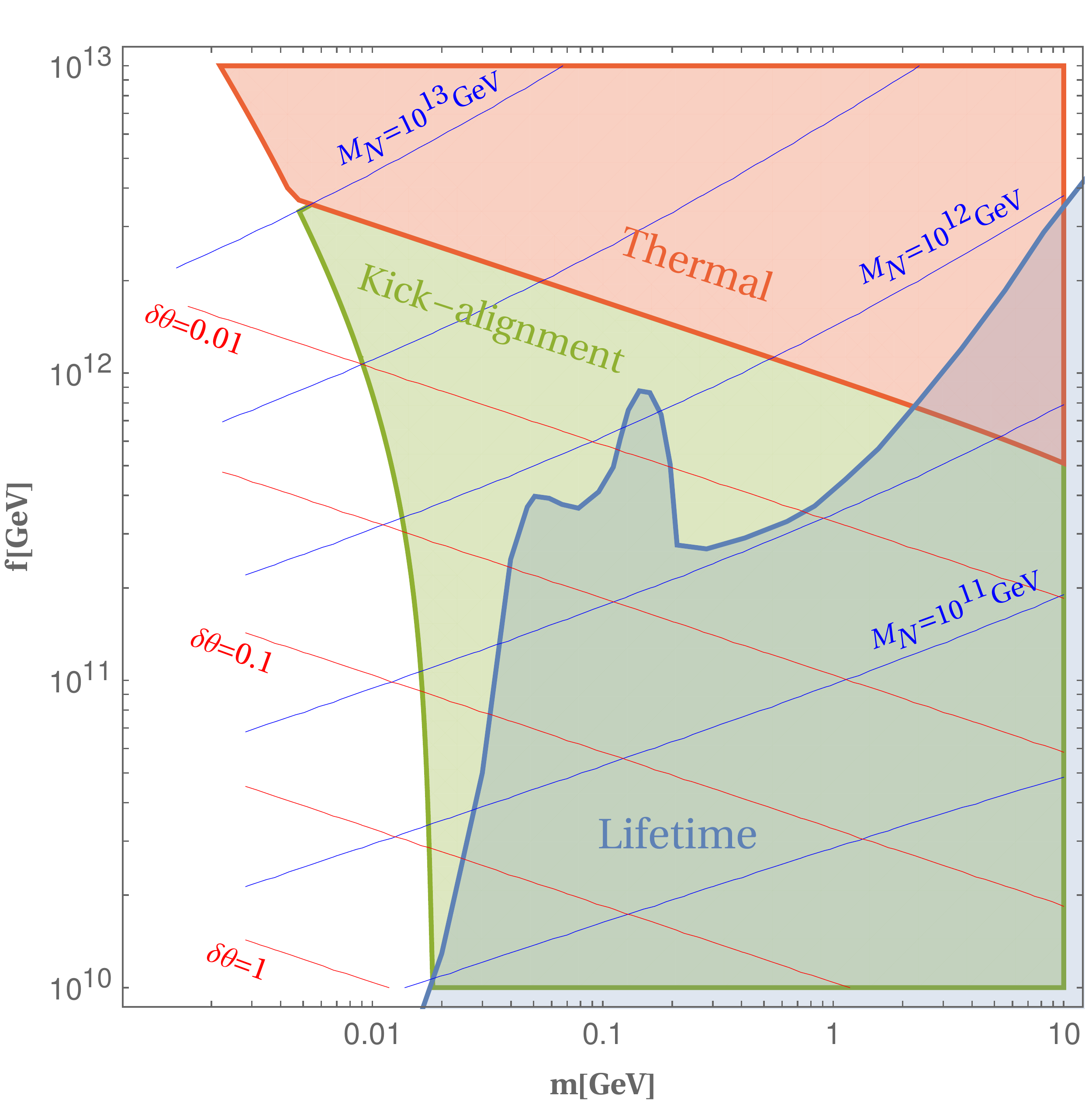}
    \caption{Leptogenesis scenario with large $K\sim M_{\rm pl} m_\nu/v^2$ and $Q=B-3L_e$, see text for details.}
    \label{fig:LG}
\end{figure}

The lifetime relative to the age of the universe in terms of this effective mass
\begin{align}
    %\frac{\phi}{\Lambda}\partial_\mu J_L^\mu&=
 %  & \frac{\phi}{f} m_\nu \nu \nu \\ 
   \frac{\tau_U}{\tau_\phi}&=0.33 \frac{(m_\nu^{\rm{eff}})^2}{\Delta m_{\rm atm}^2}\left(\frac{10^{10}{\rm GeV}}{f}\right)^2\frac{m}{\rm GeV}. 
\end{align}
This ratio has to be smaller than one by at least some five orders of magnitude~\cite{PalomaresRuiz:2007ry} which imposes a lower bound on $f$ shown as the blue-shaded region in fig.~\ref{fig:LG}.
In addition for Kick-alignment production to dominate over thermal production there is an upper bound on $f$, similarly to the GUT baryogenesis scenario. This follows, from the Kick-alignment relic abundance scaling with heavy mass as $ (M_N/f)^2\sqrt{m/{\rm GeV}}$ whereas the thermal contribution substituting $y_N^2 \sim M_N m_{\nu}^{\rm eff} $ scales then with $M_N^3$. For fixed $m$ larger $f$, the model requires larger $M_N$ and thermal production overtakes Kick-alignment in the red-shaded area of fig.~\ref{fig:LG}. In addition the already mentioned consistency condition $f>T_Q$ sets a lower bound on the mass $m$ if $\phi$ is to be the dark matter. Lines for constant shift and mass $M_N$ are shown in red and blue and extend to low masses up to $\pi f/2>T_Q$. These constraints are shown to narrow down the allowed parameter space in fig~\ref{fig:LG} to a region around $m\sim 100$~MeV, $f\sim 3\times 10^{11}$~GeV.

Finally, some notes on the detection of such candidate. A distinctive signal would be present in $B$ or $L$ violating interactions and decays. The strong bounds on lifetime  nevertheless make these events rare and hard to capture in say Hyper-Kamiokande where we can estimate that in the next 30 years the probability to see one such decay to be at the level of $10^{-3}$. 

The rarity of such processes can be understood because the bounds on lifetime of $\phi$ come from processes in which $\phi$ decays into neutrinos, anywhere in the , which are subsequently detected on Earth. On the other hand, in order to obtain the smoking gun signal, we need to observe an event in which $\phi$ decays in the detector itself -- a measurement that suffers from ratio of volume of the detector to the volume of the local dark matter density.

On the other hand, the resemblance of a Majoron means that phenomenological searches for this particle would apply to our candidate as well see e.g.~\cite{Chacko:2018uke}.
We leave the exploration of other possibilities for its detection like studying the effects on matter of a $\phi$ background to future study. 

\section{Conclusions}\label{sec:conclusion}
This work laid out the possibility of primordial plasma kicking the Goldstone boson of a SM-preserved global symmetry as a mechanism to produce dark matter density. 

We have shown that the generation of the matter-antimatter asymmetry acts as a source in the evolution of the Goldstone field. This contribution produces a displacement in the field at early times which sets the amplitude for oscillations when the zero mode enters the horizon and behaves like cold matter. The mechanism can then be described as dynamically setting the initial conditions for misalignment dark-matter production which nonetheless yields a qualitatively different prediction for relic abundance similar to how freeze-in differs from freeze-out.

This production mechanism was studied in conjunction with different baryo/leptogenesis models and we find it is a feasible possibility for dark matter production. In particular, combination with Leptogenesis seems to offer viable parameter space for dark matter production with mass $10$~MeV -- $1$~GeV and decay constant $f$ in the range of $10^{10} - 10^{12}$~GeV.

\begin{acknowledgments}
JS is very grateful for the support from the COFUND Fellowship and for the support from the research grant TAsP (Theoretical Astroparticle Physics) funded by Istituto Nazionale di Fisica Nucleare (INFN).
\end{acknowledgments}

%\appendix
\appendix

\section{Matching after the kick}
\label{sec:matching}
After Q-number generation the source term has been turned off and a homogeneous solution describes the evolution. One can then take $\phi(t_F>t_Q+\tau)$ 
\begin{align}
    \phi(t_F)=\bar C_1 \frac{J_{1/4}(m t_F)}{(mt_F)^{1/4}}+\bar C_2 \frac{Y_{1/4}(m t_F)}{(mt_F)^{1/4}}
\end{align}
and match with initial condition equation (\ref{SolEOM}) and it's derivative evaluated at $t_F$, assuming $mt_F\ll 1$ this returns
\begin{align}\label{IC2}
    \bar C_1=& 2^{1/4} \Gamma_{5/4}\Bigg(\phi(t_F)\\ \nonumber
    &\qquad\quad+2t_F\dot\phi(t_F)\left(1+\frac{\Gamma_{-1/4}\sqrt{mt_F}}{2\Gamma_{1/4}}\right)\Bigg)+\mathcal O(m t_F)\\
    \bar C_2&=\frac{2\pi (mt_F)^{1/2}}{2^{1/4}\Gamma(1/4)} t_F\dot\phi(t_F)+\mathcal O(m t_j)
\end{align}
Where $\phi$ and $t_F\dot\phi$ are comparable.
The late time behaviour reads instead $t\gg m^{-1}$:
\begin{align}
    \phi(t)\simeq \sqrt{\bar C_1^2+\bar C_2^2}\sqrt{\frac{2}{\pi(mt)^3}}\cos(mt+\omega)
\end{align}
so given the hierarchy $\bar C_1\gg \bar C_2$ and in terms of $t_m$ one obtains relation in eq.~(\ref{Late}).
In particular the 0th and first correction $\mathcal O(\sqrt{m t_F})$ come from $\bar C_1$ (from the cross term in the square as opposed to $C_2^2$ which starts at order $mt_j$) which can be  written to first order in a $t_F$ independent-form as
\begin{align}\nonumber
\int_{t_Q}^{t_m^*}\frac{n_{Q}}{f}dt'&=\int_{t_Q}^{t_F} \frac{n_{Q}(t)}{f}dt'+ \frac{n_{ Q}(t_F)}{f}\int_{t_F}^{t_m^*} \left(\frac{t_F}{t}\right)^{3/2}dt'\\
&=\phi(t_F)+2t_F\dot \phi(t_F)\left(1-\frac{\sqrt{t_F}}{\sqrt{t_m^*}}\right)
\end{align}
so direct comparison with eq.~(\ref{IC2}) and the fact that $\Gamma_{-1/4}/\Gamma_{1/4}\simeq -1.35<0$ allow to write $\bar C_1$ in a $t_j$ independent form 
\begin{align}
 \bar C_1&=2^{1/4} \Gamma_{5/4} \int_{t_Q}^{t^*_m}\frac{n_{\Delta Q}(t)}{f}dt+\mathcal O({mt_j}) \\  \label{tms}
 t_m^*&=\left(\frac{\Gamma_{1/4}}{\Gamma_{-1/4}}\right)^2\frac{4}{m}\simeq \frac{2.188}{m}\simeq 1.1 r_m^{-2} t_Q
 \end{align}
which is the field displacement as in eq.~(\ref{deltaphi}) for late enough $t$ or rewritten in terms of temperature in eq.~(\ref{DisPl}). 

\section{Integrals}\label{App:Int}
The integral of $Q$-number abundance can be cast into an integral over temperature and using the dimensionless variable $r=T/T_Q$ one can factor out the typical scale of the process to leave a dimensionless function of $r_\tau$, $r_m$. The dependence on $r_m$, to first order, can be captured in the limits of the integral with c.f. (\ref{tms})
\begin{align}
    r_m^*= \sqrt{\frac{t_Q}{t_m^*}}=\frac{|\Gamma(-1/4)|}{\Gamma(1/4)}\frac{\sqrt{t_Qm}}{2}\simeq 0.957 r_m
\end{align}
The integral of eq.~(\ref{IX}) reads 
\begin{align}
I=&\int^1_{r_m^*} dr \left(1-e^{-r_\tau^2(r^{-2}-1)}\right)\\\nonumber=&r_\tau e^{r_\tau^2}\left(\Gamma\left(\frac12,r_\tau^2\right)-\Gamma\left(\frac12,\frac{r_\tau^2}{(r_m^*)^2}\right)\right)\\
&-r_m^*\left(1-e^{r_\tau^2(1-(r_m^*)^{-2})}\right)
\end{align}
while that of eq.~(\ref{IS})
\begin{align}
    I=&\int_{r_m}^1 dr (1-e^{-2r_\tau^2(r^{-1}-1)})\\ \nonumber
    =&2r_\tau^2e^{2r_\tau^2}\left(\Gamma\left(0,2r_\tau^2\right)-\Gamma\left(0,\frac{2r_\tau^2}{r_m^*}\right)\right) \\&-r_m^*\left(1-e^{2r_\tau^2(1-(r_m^*)^{-1})}\right)
\end{align}
where $\Gamma(x,y)$ is the incomplete Gamma function:
\begin{align}
    &\Gamma(x,y)=\int_y^{\infty} z^{x-1} e^{-z}dz \end{align}\begin{align}\nonumber
    &\Gamma(x,y\gg1)\simeq y^{x-1} e^{-y}\qquad 
    \Gamma(x,y\ll1)\simeq \Gamma(x)-\frac{y^x}{x}.
\end{align}
%To start the appendixes, use the \verb+\appendix+ command.

\bibliography{kickalign}% Produces the bibliography via BibTeX.

\end{document}